\documentclass[doublecol,floatfix]{epl2}
\usepackage{amsmath}
\usepackage{graphicx}
\usepackage{color}

\title{Self-affirmation model for football goal distributions}
 \shorttitle{Self-affirmation model for football goal distributions}

\author{Elmar~Bittner\inst{1} \and Andreas~Nu{\rm \ss}baumer\inst{1} 
\and Wolfhard~Janke\inst{1} \and Martin~Weigel\inst{2}}
 \shortauthor{E. Bittner \etal}

\institute{
\inst{1} Institut f\"ur Theoretische Physik and Centre for 
			Theoretical Sciences (NTZ) -- Universit\"at Leipzig\\
			Postfach 100\,929, D-04009 Leipzig, Germany\\
\inst{2} Department of Mathematics and the Maxwell
  			Institute for Mathematical Sciences, Heriot-Watt University\\
         Riccarton, Edinburgh, EH14\,4AS, Scotland, UK
}


\abstract{
  Analyzing football score data with statistical techniques, we investigate how the
  highly co-operative nature of the game is reflected in averaged properties such as
  the distributions of scored goals for the home and away teams. It turns out that in
  particular the tails of the distributions are {\em not\/} well described by
  independent Bernoulli trials, but rather well modeled by negative binomial or
  generalized extreme value distributions. To understand this behavior from first
  principles, we suggest to modify the Bernoulli random process to include a simple
  component of {\em self-affirmation\/} which seems to describe the data surprisingly
  well and allows to interpret the observed deviation from Gaussian statistics.  The
  phenomenological distributions used before can be understood as special cases
  within this framework. We analyzed historical football score data from many leagues
  in Europe as well as from international tournaments and found the proposed models
  to be applicable rather universally. In particular, here we compare men's and
  women's leagues and the separate German leagues during the cold war times and find
  some remarkable differences.
}

\begin{document}

\maketitle


Football (soccer) is 
of the most popular sports world-wide, attracting millions of
spectators each year. Its popularity and economical importance also captivated
scientists from many fields, for instance in the attempt to improve the game tactics
etc. Much less effort has been devoted, it seems, to the understanding of football
(and other ball sports) from the perspective of the stochastic behavior of
co-operative ``agents'' ({\it i.e.}, players) in abstract models. Such problems recently
have come into the focus of physicists in the hope that the model-based point-of-view
and methodological machinery of statistical mechanics might add a new perspective to
the much more detailed investigations of more specific disciplines
\cite{stauffer:03,book0}. 
Some reports of such research are collected in Ref.~\cite{all}.
Score distributions of ball games have been occasionally
considered by statisticians \cite{moroney:56,wessons,reep:71,green}. Very small data sets
were initially found to be reasonably well described by the simplest Poissonian model
resulting from constant and independent scoring probabilities \cite{moroney:56}.
Including more data, however, better phenomenological fits were achieved with
models such as the negative binomial distribution (NBD), which can be constructed 
from a mixture of independent Poissonian processes \cite{reep:71}, or even with
models of generalized extreme value (GEV) statistics \cite{green,book}, which
are particularly suited for heavy-tailed distributions. This yielded a rather
inhomogeneous and purely phenomenological picture, without offering any microscopical
justification. We argue that the crucial ingredient missed in previous studies are
the {\em correlations\/} between subsequent scoring events.

In a broader context, this problem of extremes is of obvious importance, for instance,
in actuarial mathematics and engineering, but the corresponding distributions with
fat tails also occur in many physics fields,
ranging from the statistical mechanics of regular and disordered systems
\cite{boucaud,bramwell:00,berg,bittner} over turbulence \cite{noullez:02} 
to earth quake data \cite{varotsos:05}.
In these cases often average properties were considered instead of explicit extremes, 
and the empirical occurrence of heavy-tailed distributions led to speculations about 
hidden extremal processes, most of which could not be identified, though. It was 
only realized recently that GEV distributions can also
arise naturally as the statistics of sums of {\em correlated\/} random variables
\cite{bertin}, which could explain their ubiquity in nature. 

For the specific example of
scoring in football, correlations naturally occur through processes of feedback of
scoring on both teams, and we shall see how the introduction of simple rules for the
adaptation of the success probabilities in a modified Bernoulli process upon scoring
a goal leads to systematic deviations from Gaussian statistics. We find simple models
with a single parameter of {\em self-affirmation\/} to best describe the available
data, including cases with relatively poor fits of the NBD.
The latter is shown to result from one of these models in a particular
limit, explaining the relatively good fits observed before.

To investigate the importance of correlations, we consider the distributions of goals
scored by the home and away teams in football league or cup matches. To the simplest
possible approximation, both teams have independent and constant (small)
probabilities of scoring during each appropriate time interval of the match, such
that the resulting final scores $n$ follow a Poisson distribution,
\begin{equation}
    \label{eq:poisson}
    P_{\lambda}(n) =  \frac{{\lambda}^{n}}{n!} e^{-\lambda},
\end{equation}
where $\lambda = \langle n \rangle$. Here and in the following, separate parameters
are chosen for the scores of the home and away teams.  Clearly, this is a gross
over-simplification of the situation.  Averaging over the matches during one or
several seasons, one might expect a {\em distribution\/} of scoring probabilities
$\lambda$ depending on the different skills of the teams, the lineup for the match
etc., leading to the notion of a {\em compound\/} Poisson distribution. For the
special case of $\lambda$ following a gamma distribution $f(\lambda)$,
the resulting compound distribution is a NBD
\cite{feller},
\begin{equation}
  \label{eq:neg_binomi}
   P_{r,p}(n) = \int_0^\infty\mathrm{d}\lambda\,P_\lambda(n) f(\lambda) =\frac{\Gamma(r+n)}{n!\,\Gamma(r)} p^n(1-p)^r.
\end{equation}
The NBD form has been found to describe football score data rather well
\cite{reep:71,green}. It appears rather {\em ad hoc\/}, however, to assume that
$f(\lambda)$ follows a gamma form, and fitting different seasons of our data with the
Poissonian model \eqref{eq:poisson}, the resulting distribution of $\lambda$ does not
resemble a gamma distribution. 
As a phenomenological alternative to the NBD, Greenhough {\em et al.\/} \cite{green}
considered fits of the GEV distributions 
\begin{equation}
  \label{eq:gev}
  P_{\xi,\mu,\sigma}(n) =
  \frac{1}{\sigma}\left(1+\xi\frac{n-\mu}{\sigma}\right)^{-1-1/\xi}
  e^{-\left(1+\xi\frac{n-\mu}{\sigma}\right)^{-1/\xi}}
\end{equation}
to the data, obtaining clearly better fits than with the NBD in some cases. Depending
on the value of the parameter $\xi$, these distributions are known as Weibull ($\xi <
0$), Gumbel ($\xi \rightarrow 0$) and Fr\'echet ($\xi > 0$) distributions,
respectively \cite{book}. 

In the present context of scoring in football, goals are likely not
independent events but, instead, scoring certainly has a profound feedback on the
motivation and possibility of subsequent scoring of both teams (via direct
motivation/demotivation of the players, but also, e.g., by a strengthening of
defensive play in case of a lead). Such 
feedback can be taken into account starting from a simple Bernoulli model: consider a
match divided into, e.g., $N = 90$ time steps with both teams having the
possibility to score in each unit with a probability $p=p(n)$ depending on the number
$n$ of goals scored so far.  Several possibilities arise. For our model ``A'', upon
each goal the scoring probability is modified as $ p(n) = p(n-1) + \kappa $, with
some fixed constant $\kappa$.  Alternatively, one might consider a multiplicative
modification rule, $p(n) = \kappa p(n-1)$, which we refer to as model ``B''. 
Finally, in our model ``C'' the assumption of independence of the scoring of the two 
teams is relaxed by coupling the adaptation rules, namely by setting $p_\mathrm{h}(n) =
p_\mathrm{h}(n-1) \kappa_\mathrm{h}$, $p_\mathrm{a}(n) = p_\mathrm{a}(n-1) /
\kappa_\mathrm{a}$ upon a goal of the home (h) team, and vice versa for an away (a)
goal. If both teams have $\kappa>1$, this results in an incentive for
the scoring team and a demotivation for the opponent, but a value $\kappa <1$ is
conceivable as well.
The resulting, distinctly non-Gaussian distributions $P_N(n)$ for the total number of
goals scored by one team can be computed exactly for models ``A'' and ``B'' from a 
Pascal recurrence relation
\cite{prep},
\begin{equation}
  \label{eq:recurrence}
  P_N(n) = [1-p(n)] P_{N-1}(n) + p(n-1)P_{N-1}(n-1),
\end{equation}
where $p(n) = p_0+\kappa n$ (model ``A'') or $p(n) = p_0\kappa^n$ (model
``B''). Model ``C'' can be treated similarly \cite{prep}.

It is remarkable that this rather simple class of feedback models leads to a microscopic
interpretation of the NBD in \eqref{eq:neg_binomi} which, in fact, can
be shown to be the continuum limit of $P_N(n)$ for model ``A'', {\it i.e.},
$N\rightarrow \infty$ with $p_0 N$ and $\kappa N$ kept fixed \cite{prep}.
For the NBD parameters one finds that $r = p_0/\kappa$ and $p = 1 - \mathrm{e}^{-\kappa N}$,
such that a good fit of a NBD to the data can be understood from the effect 
of self-affirmation of the teams or players, the major ingredient of our 
microscopic models ``A'', ``B'', and ``C''.  Additionally, a certain type
of continuous microscopic model with feedback can be shown to result in a GEV
distribution \cite{bertin,prep}, such that all different types of deviations from
the Gaussian form occurring here can be understood from the correlations introduced
by feedback.

\begin{table*}[tb]
\caption{\label{tab_all}Fits and their $\chi^2$ per degree-of-freedom, $\tilde{\chi}^2=\chi^2/{\rm d.o.f.}$,
 of the phenomenological distributions
  (\ref{eq:poisson}), (\ref{eq:neg_binomi}), and (\ref{eq:gev}) as well as fits of our 
  microscopic feedback models ``A'' and ``B'' to the data
  for the East German ``Oberliga'', 
  the (West) German men's premier league ``Bundesliga'',
  the German women's premier league ``Frauen-Bundesliga''
  and the qualification stages of all past ``FIFA World Cups''.
}
\begin{center}
    \begin{tabular}{llrrrrrrrr}
         \hline \hline
         &&\multicolumn{2}{c}{Oberliga}&\multicolumn{2}{c}{Bundesliga}&\multicolumn{2}{c}{Frauen-Bundesliga}&\multicolumn{2}{c}{FIFA World Cup}\\ 
                                             \makebox[0.2cm][c]{}    &\makebox[0.2cm][c]{}
													     &\makebox[1cm][c]{Home}&\makebox[1cm][c]{Away}
                                            &\makebox[1cm][c]{Home}&\makebox[1cm][c]{Away}
                                            &\makebox[1cm][c]{Home}&\makebox[1cm][c]{Away}
                                            &\makebox[1cm][c]{Home}&\makebox[1cm][c]{Away}\\ \hline
Poisson    &$\lambda\!$    & $ 1.85(2) $ & $1.05(1)$ &  $ 1.91(1)$  & $1.16(1)$  & $ 1.78(4)$   & $1.36(4)$              & $1.53(2)$    & $0.89(1)$ \\ \hline
           &$\tilde{\chi}^2\!$& $ 12.5  $  & $ 12.8$         &  $ 9.21 $  & $ 9.13$          & $14.6 $   & $14.4$  & $18.6$             & $25.0$          \\ \hline \hline
NBD        &$p\!$          & $ 0.17(1) $ & $0.14(1)$  & $ 0.11(1)    $  & $0.09(1)$      & $ 0.45(3) $  & $0.46(3) $     & $0.37(2)$     & $0.38(2)$ \\
           &$r\!$          & $ 9.06(88)$ & $6.90(84)$ & $ 16.2(1.9)$  & $12.1(1.7)$  & $ 2.38(24)$  & $1.97(22)$     & $3.04(21)$    & $1.76(12)$ \\ \hline
           &$p_0\!$        & $ 0.0191$   & $0.0112$   & $ 0.0202$       & $0.0125$       & $0.0160$     & $0.0133$       & $0.0154$           & $0.0094$ \\
           &$\kappa\!$     & $ 0.0021$   & $0.0016$   & $ 0.0012$       & $0.0010$       & $0.0067$     & $0.0068$       & $0.0051$           & $0.0053$ \\\hline
           &$\tilde{\chi}^2\!$& $ 0.99  $  & $ 4.09$   & $ 1.08$   & $2.22$         & $ 2.32$      & $ 1.37$       & $2.67$             & $2.02$ \\ \hline \hline  
GEV        &$\xi\!$        & $ -0.05(1)$ & $0.02(1)$  & $ -0.10(1)$  & $-0.02(1)$ & $ 0.04(4)$  & $0.25(7)$              & $0.11(2)$    & $0.19(2)$     \\
           &$\mu\!$        & $ 1.12(2) $ & $0.49(2)$  & $ 1.17(2) $  & $0.57(1) $ & $ 0.83(8)$  & $0.77(7)$              & $0.86(3)$    & $0.36(3)$     \\
           &$\sigma\!$     & $ 1.30(2) $ & $0.90(2)$  & $ 1.33(1) $  & $0.96(1) $ & $ 1.49(6)$  & $1.18(5)$              & $1.21(3)$    & $0.86(2)$     \\ \hline
           &$\tilde{\chi}^2\!$& $1.93$  &$5.04$  & $3.43$       & $7.95$     & $ 3.40 $    & $ 1.55$               & $0.85$             & $1.89$              \\ \hline \hline
Model ``A''&$p_0\!$        &$0\!.0188(2) $ & $0.0112(1)$& $0.0199(2)$ & $0.0125(2)$ & $0.0159(5)$ & $0.0132(4)$            & $0.0152(3)$& $0.0093(2)$ \\
           &$\kappa\!$     &$0.0024(2) $ & $0.0018(2)$& $0.0015(1)$ & $0.0012(1)$ & $0.0070(5)$ & $0.0071(7)$            & $0.0053(3)$& $0.0055(3)$ \\ \hline
           &$\tilde{\chi}^2\!$& $1.07$ & $4.23$  & $1.01$      & $2.31$      & $2.28$      & $1.44$                & $2.88$             & $2.19$              \\ \hline \hline
Model ``B''&$p_0\!$         & $0.0189(2)$  & $0.0112(1)$  & $0.0200(2)$ & $0.0125(1)$ &$0.0166(5)$ & $0.0138(4)$         & $0.0155(2)  $& $0.0095(2)$ \\
           &$\kappa\!$     & $1.1115(83)$ & $1.153(15)$ & $1.0679(60)$ & $1.093(11)$ & $1.315(31)$ & $1.412(55)$ & $1.278(13)$& $1.478(35)$ \\ \hline
           &$\tilde{\chi}^2\!$&$0.75$&$3.35$      &$1.25$      &$1.96$       & $3.24$      & $0.95$                & $0.92$             & $0.80$              \\ \hline \hline
    \end{tabular}
\end{center}
\end{table*}

We now confront these models with empirical 
data sets, starting with  football matches played in German
leagues, namely the ``Bundesliga'' (men's premier league (West) Germany, 1963/1964 -- 2004/2005,
$\approx 12\,800$ matches), the ``Oberliga'' (men's premier league East Germany, 1949/1950 --
1990/1991, $\approx 7700$ matches), and the ``Frauen-Bundesliga'' (women's premier
league Germany, 1997/1998 -- 2004/2005, $\approx 1050$ matches)~\cite{daten}.
We determined histograms estimating the probability density functions (PDFs)
$P^\mathrm{h}(n_h)$ and $P^\mathrm{a}(n_a)$ of the final scores of the home and away
teams, respectively~\cite{fn1}. Error estimates on the histogram bins were computed
with the bootstrap resampling method. This allows the judgment of the quality of 
the various fits 
collected in Table~\ref{tab_all}
by monitoring their goodness or $\chi^2$ per degree-of-freedom, $\tilde{\chi}^2=\chi^2/{\rm d.o.f.}$,
naturally taking into account the different numbers of free parameters in the fits considered.

\begin{figure}[tb]
\includegraphics[scale=0.8]{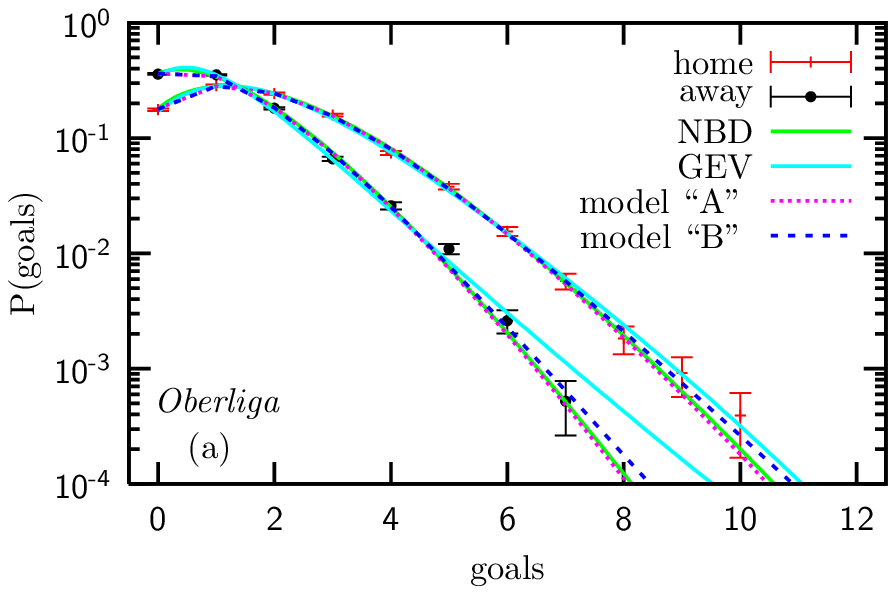}
\includegraphics[scale=0.8]{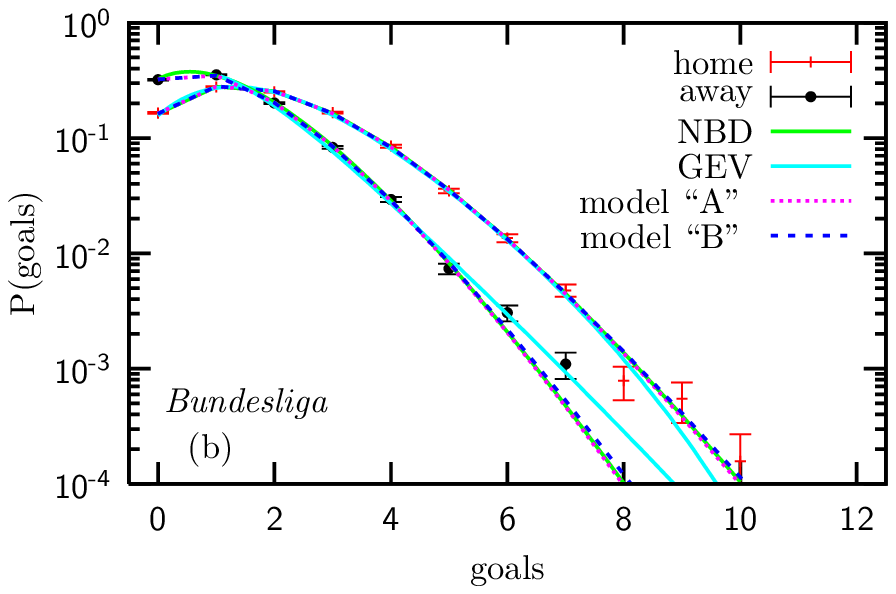}
\includegraphics[scale=0.8]{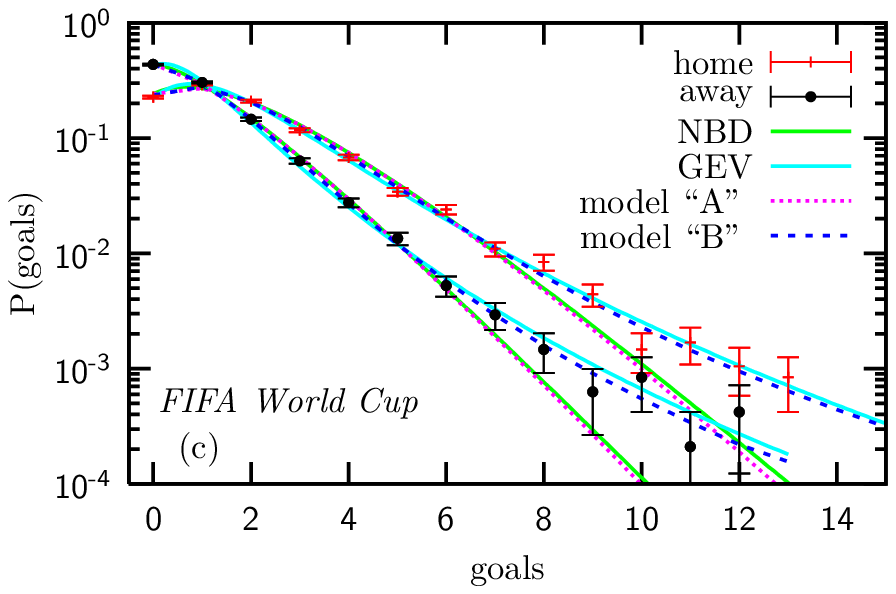}
\caption{Histograms of final scores of home and away teams and corresponding fits.
(a) East German ``Oberliga''.
(b) (West) German ``Bundesliga''.   
(c) The qualification stage of the ``FIFA World Cup'' series. 
}
\label{fig_all}
\end{figure}

We first considered fits of the PDFs of the
phenomenological descriptions (\ref{eq:poisson}), (\ref{eq:neg_binomi}), and
(\ref{eq:gev}). 
Not to our surprise, and in accordance with previous findings \cite{reep:71,green}, the
simple Poissonian ansatz (\ref{eq:poisson}) is not found to be an adequate
description for any of the data sets. Deviations occur here mainly in the tails with
large numbers of goals which in general are found to be fatter than can be
accommodated by a Poissonian model. On the contrary, the NBD form
(\ref{eq:neg_binomi}) models all of the above data well as is illustrated 
in Fig.~\ref{fig_all}.
Considering the fits of the GEV distributions (\ref{eq:gev}),
we find that extreme value statistics are in general also a reasonably good
description of the data. The shape parameter $\xi$ is always found to be small in
modulus and negative in the majority of the cases, indicating a distribution of the
Weibull type (which is in agreement with the findings of Ref.~\cite{green} for
different leagues). Fixing $\xi = 0$ yields overall clearly larger
$\chi^2$ values. 
Comparing ``Oberliga'' and ``Bundesliga'',
we consistently find larger values of the parameter $\xi$ for the former, indicative
of the comparatively fatter tails of these data, see Table~\ref{tab_all} and
Fig.~\ref{fig_all}.
Comparing to the results for the NBD, we do not find any cases where the GEV
distributions would provide the best fit to the data, so clearly the leagues
considered here are not of the type for which Greenhough {\em et al.\/}~\cite{green}
found better matches with the GEV statistics than for the NBD.  Similar conclusions
hold true for the comparison of ``Bundesliga'' and ``Frauen-Bundesliga'', with the
latter taking on the role of the ``Oberliga''.

Representing the continuum limit of our model ``A'', the good performance of the
NBD fits observed so far implies that the feedback models proposed here
can indeed capture the main characteristics of the game. To test this
conjecture directly we performed 
fits of the {\em exact\/} distributions resulting from the recurrence relation (\ref{eq:recurrence}),
employing the simplex method 
to minimize the total $\chi^2$ deviation for the home and away scores.
Comparing the results of model ``A'' to the fits of the limiting NBD, we observe
in Table~\ref{tab_all}
almost identical fit qualities for the final scores. However, for sums
and differences of scores we find a considerably better description by using our model ``A'',
indicating deviations from the continuum limit there ~\cite{prep}. 
The overall best modeling of the league
data is achieved with fits of model ``B'' which feature
on average an even higher
quality than those of model ``A'', cf.\ Table~\ref{tab_all}.
We also performed fits to the more elaborate model ``C'', but found the results
rather similar to those of the simpler model ``B'' and hence do not 
discuss them here. 

Comparing the leagues, we see in Table~\ref{tab_all} that the parameters $\kappa$ for the
``Oberliga'' are significantly larger than for the ``Bundesliga'',
whereas the parameters $p_0$ are slightly smaller for the ``Oberliga''.
That is to say, scoring a goal in a match of the East German
``Oberliga'' was a more encouraging event than in the (West) German ``Bundesliga''.
Alternatively, this observation might be interpreted as a stronger tendency of the
perhaps more professionalized teams of the (West) German premier league to switch to a strongly
defensive mode of play in case of a lead.
Consequently, the tails of the distributions are slightly fatter for the ``Oberliga''
than for the ``Bundesliga''.
Recalling that the NBD form (\ref{eq:neg_binomi}) is the
continuum limit of the feedback model ``A'', these differences should
translate into larger values of $r$ and smaller values of $p$ for the
``Bundesliga'' results, which is what we indeed observe. Conversely, computing from
the NBD parameters $r$ and $p$ the feedback parameters $p_0$ and $\kappa$ also given
in Table~\ref{tab_all}, we obtain good agreement with the directly fitted values.
Comparing the results for the ``Frauen-Bundesliga'' to those for the ``Bundesliga'',
even more pronounced tails are found for the former, resulting in very significantly
larger values of the self-affirmation parameter $\kappa$.

Finally, we also considered the score data of the qualification stage of the ``FIFA
World Cup'' series from 1930 to 2002 ($\approx 3400$ matches)
\cite{daten5,fn3}. 
Compared to the domestic league data discussed
above, the results of the World Cup show distinctly heavier tails, cf.\
Fig.~\ref{fig_all}. 
Consequently we obtain good fits for the heavy-tailed
distributions, and, in particular, in this case the GEV distribution provides a
better fit than the NBD,
similar to what was found by Greenhough {\em et al.\/}~\cite{green}, cf.\ Table~\ref{tab_all}. 
The fits of model ``A'' are again rather similar to the NBD. The 
multiplicative feedback model ``B'', on the other hand, also handles this
case extremely well and, for the away team, considerably better than the 
GEV distribution (\ref{eq:gev}).
The difference
to the league data can be attributed to the possibly very large differences in skill
between the opposing teams occurring since all countries are allowed to participate
in the qualification round. The parameters in Table~\ref{tab_all} reveal a remarkable
similarity with the parameters of the ``Frauen-Bundesliga'', where a similar
explanation appears quite plausible since the very good players are concentrated in
just two or three teams. 

We have shown that football score data can be understood from a certain class of
modified binomial models with a built-in effect of self-affirmation of the teams upon
scoring a goal. The NBD fitting many of the data sets can in fact be understood as a
limiting distribution of our model ``A'' with an additive update rule of the
scoring probability.
It does not provide very good fits in cases with heavier tails, such as the
qualification round of the ``FIFA World Cup'' series. The overall best variant is our
model ``B'', where a multiplicative update rule ensures that each goal motivates the
team even more than the previous one. Basically by ``interpolating'' between the GEV 
form and NBD, it fits both these world-cup data as well as the data
from the German domestic leagues extremely well, thus reconciling the heterogeneous
phenomenological findings with a plausible and simple microscopic model.
In general, we find less professionalized leagues or cups to feature stronger scoring
feedback, resulting in goal distributions with heavier tails. It is obvious that the
presented models with a single parameter of self-affirmation are a bold
simplification. It is all the more surprising then, how rather well they model the
considered score data, yielding a new example of how sums of correlated variables
lead to non-Gaussian distributions with fat tails. For a closer understanding of the
self-affirmation effect, an analysis of time-resolved scoring data would be highly
desirable. Some data of this type has been analyzed in Ref.~\cite{dixon}, showing a
clear increase of scoring frequency as the match progresses, thus supporting the
presence of feedback as discussed here.

\acknowledgments
  The authors are grateful to O.\ Penrose and S.\ Zachary for discussions.  This work
  was partially supported by the DFG under grant Nos.\ JA483/22-1 \& 23-1 and the EC
  RTN-Network `ENRAGE': {\em Random Geometry
  and Random Matrices: From Quantum Gravity to Econophysics\/}
  under grant
  No.\ MRTN-CT-2004-005616.
  M.W.\ acknowledges support by the EC MC-EIF programme under contract No.\
  MEIF-CT-2004-501422.


\begin{thebibliography}{99}
\bibitem{fn3}
We disregarded all games played in tournaments on neutral grounds. The final knockout
stage follows different rules \cite{prep}.
\end{thebibliography}
\end{document}